# Implementing a Language for Distributed Systems

## Choices and Experiences with Type Level and Macro Programming in Scala

Pascal Weisenburger[a] and Guido Salvaneschi[a]

a  Technische Universität Darmstadt, Germany

**Abstract**   Multitier programming languages reduce the complexity of developing distributed software by developing the distributed system within a single coherent code base. In multitier languages, the compiler or the runtime takes care of separating the code into the components of the distributed system. This approach enables abstraction over low level implementation details such as data representation, serialization and network protocols. The ScalaLoci programming language allows developers to declare the components of the system and their architectural relation at the type level, enabling static reasoning about distribution and remote communication and guaranteeing static type safety for data transfer across components. As the compiler automatically generates the communication boilerplate among components, data transfer among components can be modeled declaratively, by specifying the data flows in the reactive programming style.

In this paper, we report on the ScalaLoci implementation and on our experience with embedding ScalaLoci's language features into Scala as a host language. We show how a combination of Scala's advanced type level programming and of Scala's macro system enable enriching the language with new abstractions for distributed systems. We describe the challenges we encountered for the implementation and report on the solutions we developed. Finally, we outline suggestions for improving the Scala macro system to better support embedding domain-specific abstractions.

# The Art, Science, and Engineering of Programming

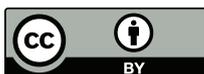

**Implementing a Language for Distributed Systems**

## 1 Introduction

Developing distributed systems adds significant complexity over programming local applications. Many issues that arise in the distributed setting are not specifically supported by existing programming languages, including, for example, consistency, availability through data replication, and fault tolerance. More generally, widely used languages provide poor abstractions for distribution, either (i) making distribution completely transparent (e.g., distributed shared memory [38]), (ii) providing a high-level programming model only for a specific domain, such as big data or streaming systems [16, 62, 10] or (iii) being general purpose but providing only low-level communication primitives, such as message passing in the actor model [1]. *Multitier* (sometimes called *tierless*) languages provide a holistic view on the distributed application, mixing functionalities belonging to different tiers (e.g., the client and the server) within the same compilation unit [14, 53]. The ScalaLoci[1] [60, 61] multitier language supports (i) declaratively specifying the distributed components and their architectural relation and (ii) placing values on the specified components. Remote communication is statically checked for architectural conformance and type safety.

In this paper, we present the implementation of ScalaLoci. We provide insights into our approach of embedding ScalaLoci abstractions as a domain-specific language into Scala and describe our experiences with using Scala's type level programming features and Scala's macro system to perform extensive AST transformations. To the best of our knowledge, the ScalaLoci project accounts for the most extensive use of Scala macros to date (with 5.5 K lines of code for macro expansion, compared to ∼ 3.5 K for *Scala Spores* mobile closures [34] and the *ScalaTest* [59] testing framework, ∼ 2 K for *Scala Async* asynchronous programming abstractions [21], ∼ 2 K for the *shapless* generic programming library [48], and ∼ 1 K for the *circe* JSON library [6]). Our work on ScalaLoci is also an experiment on Scala's expressiveness in terms of type level programming, macro programming and syntactic flexibility. In summary, this paper makes the following contributions:

- We provide an overview of the ScalaLoci design requirements and of the ScalaLoci implementation.
- We show how the ScalaLoci design is embedded into Scala and how ScalaLoci's domain-specific abstractions are translated into plain Scala with a combination of type level and macro programming.
- We present the design of our network communication runtime which hides the implementation details of different underlying network protocols and the semantics of accessing values of different types.

Section 2 lays out the design space and gives a high-level introduction to the ScalaLoci language. Section 3 provides an overview of the ScalaLoci implementation. Section 4 presents the type level encoding of the language constructs. Section 5 reports on our macro-based code splitting implementation. Section 6 describes the communication runtime. Section 7 discusses related work. Section 8 concludes.

---

[1] http://scala-loci.github.io

## 2 Design Approach of ScalaLoci

The design of ScalaLoci unfolds around three major ideas. First, the distributed topology (i.e., separate locations that execute code) is explicit and specified by the programmer, who assigns code to each system component. Second, remote communication within the distributed system (i.e., with performance and failure characteristics different from local communication) is explicit and supports event-based interaction between components. Third, ScalaLoci abstractions are embedded as domain-specific features into an existing general purpose language. We lay out the design space for our language implementation and derive concrete design requirements.

**Explicit Specification of Distribution and Topology**   Some languages for distributed systems abstract over the topology, i.e., code is agnostic to the system's components and their connections. Such approach is often chosen for highly specialized programming models, e.g., streaming systems [16, 62], intentionally hiding distribution. On the other end lie approaches where developers specify the topology. Distribution unavoidably becomes apparent when implementing different components, e.g., in actor systems [1]. While actor systems encapsulate components into actors, topological information is not encoded at the language level but managed explicitly by the developer.

Clearly, a multitier programming model for developing generic distributed systems – similar to what the actor model offers – cannot abstract over distribution completely. To decide on the component that executes a specific part of the code, the compiler or the runtime may determine the location automatically by splitting code, i.e., the developer gives up control over where code is executed. Alternatively, developers can annotate parts of the code with the location where it should be executed.

We decided for an annotation-based approach that gives the developer full control over the distribution, making locations explicit in the language. We think that it is essential for the developer to restrict the locations where code is executed for comprehending the system's behavior, e.g., regarding performance or privacy. Mixing automatic and explicit placement of values remains to be explored further.

Encoding the topology statically allows for providing static guarantees, such as that remote communication adheres to the topology and is statically type-checked. Thus, we (i) make location annotations part of a value's type and (ii) encode the topology specification at the type level, which allows the compiler to check correctness of remote accesses and ensures type safety across distributed components.

**Explicit Remote Access and Event-based Communication**   Communication in distributed systems often follows a message passing approach (such as in actor systems). Message passing mechanisms, however, are quite low level with explicit send and receive operations [20], which disrupt control flow between a sending and a receiving side. Further, message loops often need to pattern-match on received messages since messages are essentially untyped. Instead, we favor strongly typed communication mechanisms.

Message passing semantics is close to sending packets over the network whereas communication mechanisms like remote procedure calls are closer to abstractions commonly found in high-level languages. For remote procedures, however, it is impor-tant to consider the inherent differences between remote calls and local calls [26]. In contrast to local calls, remote calls need to account for network latency and potential communication failures. For this reason, in our design, remote calls return a value that represents an asynchronous computation which may potentially fail (e. g., a future).

Further, distributed applications are often event-based [11, 31]. For example, web clients send events to the server to trigger computations or cause a change to persistent storage. Servers send events to reflect such changes on the clients. Hence, we support specifying data flow in a declarative way using reactive programming features [51, 49].

**Embedding** There is a trade-off between designing a new language from scratch, which provides the greatest amount of flexibility, and embedding new abstractions into an existing language, which restricts the design space since the host language dictates the underlying feature set and available syntax. On the other hand, embeddings enable leveraging existing libraries and tooling.

We decided to embed our abstractions into Scala for its powerful type system, which can encode the topology specification and the placement of values. Our approach retains compatibility with plain Scala and preserves access to the Scala (and Scala.js) ecosystem. We exclusively use syntactically valid and type-correct Scala, allowing for expressing the placement of values in their types. The special semantics of placed values is implemented by a compile-time transformation based on macro expansion.

## 2.1 Design Requirements

From the design considerations above, we derive the following principles to guide our language implementation:

**#1 Support different architectural models** Distributed systems exhibit different architectures. Besides common schemes like client–server or a peer-to-peer, developers should be able to freely define the distributed architecture declaratively (section 4).

**#2 Make remote communication direct and explicit** The programmer should be able to execute a remote access by directly accessing values placed on a remote peer. Although remote communication boilerplate code should be reduced to a minimum, remote access should still be syntactically noticeable since it entails potentially costly network communication (section 4.1).

**#3 Provide static guarantees** The language should provide static guarantees whenever possible to catch errors preferably already at compile-time. In particular, access to remote values should be statically checked to conform to the specified distributed architecture and to provide type safety across distributed components (section 4.2).

**#4 Support declarative cross-host data flows** The language should support abstractions for reactive programming for specifying data flows across hosts since distributed applications are often reactive in nature (section 6).

**#5 Make abstractions for distribution integrate with existing code** The new language abstractions for distributed programming should be orthogonal to present language abstractions and integrate properly with existing code. Embedding our abstractions into a host language fosters reusability of existing code (section 5).

## 2.2 ScalaLoci in Nuce

The ScalaLoci multitier language supports generic distributed architectures. Developers can freely define the different components, called *peers*, of the distributed system and their architectural relation. The application-level network topology in ScalaLoci is encoded at the type level. Peers are defined in Scala as abstract type members:

```
1  @peer type Registry
2  @peer type Node
```

In Scala, traits, classes and objects can define type members, which are either abstract (e.g., type T) or define concrete type aliases (e.g., type T = Int). Abstract types can define lower and upper type bounds (e.g., type T >: LowerBound <: UpperBound), which refine the type while keeping it abstract. We use peer types only as *phantom types* [12] to keep track of placement at the type level. Hence, they are never instantiated and, thus, are defined abstract. We use Scala annotations (i.e., @peer) to distinguish peer types from other type member definitions.

We further use peer types to express the architectural relation between the different peers by specifying *ties* between peers. Ties statically approximate the run time connections between peers. For example, a tie from a Registry peer to a Node peer defines that, at run time, Registry instances can connect to Node instances. To give remote accesses useful static types, ties differentiate between different multiplicities. A *single* tie expresses the expectation that a single remote instance is always accessible. When accessing a remote value over single tie, the single value is transmitted to the local instance. An *optional* tie allows at most one remote instance to be connected and remote access locally creates a value of an option type. A *multiple* tie allows an arbitrary number of connected remote instances and remote access locally creates a sequence containing the remote values for all connected instances.

Remote access is statically checked against the architectural scheme specified through ties. Hence, ties are also encoded at the type level such that that compiler can check that the code conforms to the specified architecture. Ties are defined by specifying a type refinement for peer types that declares a Tie type member:

```
1  @peer type Registry <: { type Tie <: Multiple[Node] }
2  @peer type Node <: { type Tie <: Single[Registry] with Multiple[Node] }
```

The type refinement { type Tie <: Multiple[Node] } specified as upper bound for Registry states that Registry is a subtype of a type that structurally contains the definition of the Tie type member type Tie <: Multiple[Node]. The tie specification above defines (i) a multiple tie from the Registry to the Node peer, i.e., the registry can connect to multiple nodes, and (ii) a single tie from the Node to the Registry peer as well as a multiple tie from the Node to the Node peer, i.e., a node always connects to a single registry and can connect to multiple other nodes.

Having defined the components and their architectural relation using peer types, developers can place values on the different peers through *placement types* [60]. The placement type `T on P`[2] represents a value of type `T` on a peer `P`. Placed values are initialized by `placed { e }` expressions. The snippet places an integer on the registry:

```scala
val i: Int on Registry = placed { 42 }
```

Accessing remote values requires the `asLocal` marker, creating a local representation of the remote value by transmitting it over the network:

```scala
val j: Future[Int] on Node = placed { i.asLocal }
```

Calling `i.asLocal` returns a future of type `Future[Int]`, accounting for network delay and potential communication failure. Futures – which are part of Scala's standard library – represent values that will become available in the future or produce an error.

ScalaLoci multitier code resides in *multitier modules,* i.e., in classes, traits or objects that carry the `@multitier` annotation. Multitier modules can be combined using Scala's mixin composition on traits or by referencing instances of multitier modules [61]. Considering the following architecture specification for a module which provides a peer that monitors other peers (e.g., using a heartbeat mechanism):

```scala
@multitier trait Monitoring {
  @peer type Monitor <: { type Tie <: Multiple[Monitored] }
  @peer type Monitored <: { type Tie <: Single[Monitor] }
  // …
}
```

The following module reuses the monitoring functionality by defining an `object` extending the `Monitoring` trait to instantiate the `Monitoring` module (line 2):

```scala
@multitier trait P2P {
  @multitier object mon extends Monitoring

  @peer type Registry <: mon.Monitor {
    type Tie <: Multiple[mon.Monitored] with Multiple[Node] }
  @peer type Node <: mon.Monitored {
    type Tie <: Single[mon.Monitor] with Single[Registry] with Multiple[Node] }
  // …
}
```

The P2P module defines the `Registry` peer to be a special `Monitor` peer (line 4) and the `Node` peer to be a special `Monitored` peer (line 6) by declaring a subtype relation (e.g., `Registry <: mon.Monitor`) to map the architecture of the `Monitoring` module to the architecture of the P2P module, reusing the monitoring functionality. Refining the upper bound of a peer type enables specializing a peer as (a subtype of) another peer, enabling peer composition by combining super peers into a sub-peer. By defining that a `Registry` peer *is a* `Monitor` peer, all values placed on `Monitor` are also available on `Registry`. We use the path-dependent type `mon.Monitor` to refer to the `Monitor` peer of the multitier module instance `mon`. Scala types can be dependent on an path (of objects). Hence, we can distinguish between the peer types of different multitier module instances, i.e., the type members defined in different Scala objects.

---

[2] `T on P` is infix notation for the parameterized type `on[T, P]`.

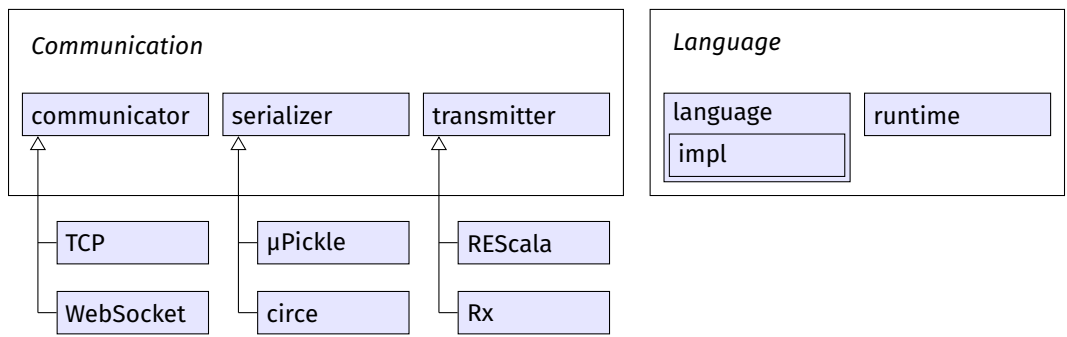

**Figure 1** Implementation Overview

## 3 Overview of the ScalaLoci Architecture

Figure 1 provides an overview of the ScalaLoci implementation, which is organized into two projects. The *communication* project (section 6) handles network communication for the generated peer-specific code of the multitier program. The project is divided into three packages that provide interfaces for type-safe remote communication over different network protocols. The remote communication mechanisms are not hardwired into our implementation and can be extended by implementing the interfaces for (i) message passing over different underlying network protocols (*communicators*), (ii) different data serialization schemes (*serializers*) and (iii) different transmission semantics (*transmitters*), e.g., pull-based remote procedure calls and push-based event streams. We implemented support for different network protocols (e.g., TCP and WebSocket), serializers (e.g., using µPickle [27] or circe [5] for serialization to JSON) and reactive systems (e.g., REScala [50] and Rx [32]). Developers can plug in such implementations as needed for configuring remote communication.

The *language* project provides the runtime package implementing the peer instance life cycle of starting and stopping peers and dispatching remote accesses using the communication backend. The language package contains the encoding of our domain-specific abstractions into Scala and the Scala type system (section 4). The language.impl package implements the macro-based code generation (section 5).

### 3.1 Cross-Platform Compilation

We support both the standard Scala compiler emitting Java bytecode and the compilation to JavaScript using the Scala.js compiler. Since there exist libraries which are available only for the JVM or only for JS, multitier modules can mix JVM-only and JS-only libraries if some peers are supposed to run on the JVM and some on a JS virtual machine. Our approach requires a multitier module to be compiled once for every platform, on which one of its peers runs. While the implementation of some libraries may not be available to all platforms, the typing information of their definitions is, i.e., the JS compiler can type-check code even if it refers to JVM-only libraries and vice versa. After type-checking and splitting the code, Scala.js' dead code elimination removes all references to JVM libraries, which are not invoked from JavaScript.

## 4 Type Level Encoding

ScalaLoci features the specification of the distributed architecture at the type level by defining (i) the different components as peers and (ii) the topology in which peers are arranged as ties (cf. design principle #1). The encoding solely relies on a combination of standard Scala features (Scala annotations, abstract type members and type refinements). For a peer definition @peer type Registry <: { type Tie <: Multiple[Node] }, we use valid Scala code instead of deviating more radically from the Scala syntax to provide a more uncluttered definition (e. g., peer Registry ties Multiple[Node]). Using standard Scala allows developers to define peers in a syntax with which they are familiar, keeping the appearance of the host language for the domain-specific aspects. Values can be placed on the components defined in the architecture specification and remote access to such values is statically checked by the compiler.

### 4.1 Lexical Context

The example in section 2.2 defines an integer value on the registry and a remote access to the value on the node using asLocal. Requiring the asLocal syntactic marker makes remote access explicit (cf. design principle #2):

```
1  val i: Int on Registry = placed { 42 }
2  val j: Future[Int] on Node = placed { i.asLocal }
```

Calling i.asLocal returns a future of type Future[Int]. We know from the tie specification that there is a single tie from Node to Registry. Thus, asLocal returns a single future. For an optional tie, the asLocal call would return an optional future of type Option[Future[Int]]. For a multiple tie, asLocal would return a sequence of futures.

The example demonstrates the interplay of placement types, peer types and ties when type-checking code (cf. design principle #3). Since the remote access i.asLocal happens on the Node peer, the value i is placed on the Registry peer and Node defines a single tie to Registry, the type system can infer the type of i.asLocal to be Future[Int].

When type-checking the asLocal call, it is necessary to determine on which peer the call is invoked. For instance, the exact same invocation on the registry peer does not type-check since there is no tie from the Registry to the Registry:

```
1  val j: Future[Int] on Registry = placed { i.asLocal } // ✗ compilation error
```

Thus, typing remote accesses depends on their lexical context. Context information in Scala can be propagated implicitly using implicit values as arguments. The placed expression desugars to a function that takes an implicit argument for the peer context:

```
1  val j: Future[Int] on Node = placed { implicit ctx: Placement.Context[Node] => i.asLocal }
```

When type-checking the i.asLocal expression, the compiler resolves a value of type Placement.Context from the implicit scope, thereby inferring its type parameter which statically captures the peer context. Using implicit functions to propagate context is a common pattern in Scala. Language support for contextual abstractions [39] will be part of Scala 3, allowing for syntactically more lightweight context propagation by omitting the definition of the implicit argument (i. e., implicit ctx: Placement.Context[Node]).

Building on Scala 2, which does not support such abstractions yet, we use Scala's macro system to synthesize implicit arguments before type-checking (section 5.4).

### 4.2 Distributed Architecture

Type-checking remote accesses heavily relies on Scala's type level programming features involving implicit resolution. The interface for accessing a remote value can be freely defined using an *implicit class*, a mechanism to define extension methods for an already defined type. The following code shows the declaration of the asLocal method used in the previous examples for placed values of type V on R accessed from a local peer L resulting in a local value T for single ties:

```scala
implicit class BasicSingleAccessor[V, R, T, L](value: V on R)(
    implicit ev: Transmission[V, R, T, L, Single]) {
  def asLocal: T = // ...
}
```

The implementation requires an implicit value of type Transmission (line 2). The implicit Transmission value again requires implicit resolution for several values to resolve (i) the current peer context as parameter L, (ii) the tie multiplicity from L to R and (iii) the type of the local representation of V as parameter T.

The resolution for the current peer context L requires an implicit argument of type Placement.Context[L] (section 4.1). After resolving the current peer context L – and knowing the peer R on which the accessed value is placed by its type V on R – the tie from L to R can be resolved using the following scheme:

```scala
sealed trait Tie[L, R, M]

sealed trait TieMultiple { /* ... */ }

sealed trait TieOptional extends TieMultiple { /* ... */ }

sealed trait TieSingle extends TieOptional {
  implicit def single[L, R](
      implicit ev: L <:< Any { type Tie <: Single[R] }): Tie[L, R, Tie.Single]
}

object Tie extends TieSingle {
  type Single
  type Optional
  type Multiple
}
```

The single method (line 8) resolves a Tie[L, R, Tie.Single] specifying a single tie from L to R. The method implicitly requires a *generalized type constraint* <:< (line 9) which the compiler resolves if L is a subtype of Any { type Tie <: Single[R] }, i.e., if the current peer has a Tie that is a subtype of Single[R]. The resolution for optional and multiple ties is defined analogously (lines 3 and 5) – left out for brevity. Letting TieSingle inherit from TieOptional and TieOptional from TieMultiple prioritizes the resolution of single ties over optional ties over multiple ties. If no tie can be resolved, peer L is not tied to peer

R and remote access is not possible. It is necessary to find suitable formulation for determining ties that can be resolved by the Scala compiler since type inference is not specified and implicit search is not guaranteed to be complete. In practice, finding such an encoding requires trying out different formulations.

The type of the local representation T usually resolves to Future[V], wrapping the accessed value into a future to take network transmission into account. Depending on the concrete data type T, other local representations that are more appropriate may be defined. For example, a remote event stream Event[T] can be locally represented simply as an Event[T] (instead of a Future[Event[T]]), which starts propagating events upon remote access. Based on the type, the compiler resolves a suitable transmission mechanism from the implicit scope. The resolved transmission mechanism connects the language level to the communication runtime (section 6).

In a similar way, variants of asLocal for optional and multiple ties return an optional value and a sequence, respectively. Note that we call the asLocal variant for accessing remote values on a multiple tie asLocalFromAll to make the cost of accessing potentially multiple remote instances visible:

```scala
implicit class BasicOptionalAccessor[V, R, T, L](value: V on R)(
    implicit ev: Transmission[V, R, T, L, Optional]) {
  def asLocal: Option[T] = // ...
}

implicit class BasicMultipleAccessor[V, R, T, L](value: V on R)(
    implicit ev: Transmission[V, R, T, L, Multiple]) {
  def asLocalFromAll: Seq[(Remote[R], T)] = // ...
}
```

Encoding the distributed architecture at the type level enables leveraging Scala's expressive type level programming features to type-check remote access based on the type of the accessed value and the architectural relation between the accessing and the accessed peer, guaranteeing static type safety across components.

### 4.3 Lessons Learned

The type level encoding currently adopted in ScalaLoci is a revised version based on our experiences with our prior implementation. Initially, we defined placement types T on P as trait on[T, P]. Depending on the implicit peer context, our implementation provided an implicit conversion T on P => T for local access and T on P => BasicSingleAccessor for remote access on a single tie (analogously for optional and multiple ties). We (i) introduced the approach using implicit classes (section 4.2) instead of using an implicit conversion for remote access and (ii) defined placed types as type alias type on[T, P] = Placed[T, P] with T, i.e., local access does not require an implicit conversion since the compound type directly entails the local representation T (and a placement marker Placed[T, P]). We decided to remove the need for implicit conversions from our encoding since implicit conversions are only applied if the compiler can infer the target type of the conversion and the compiler does not chain different implicit conversions automatically. Further, reducing the amount of required implicit search improves compile times. The downside of the revised encoding is that a placed value can always be accessed as a local value – even if it is placed on a remote peer. We can, however, reject such illegal access using a check during macro expansion.

Our domain-specific embedding into Scala is designed to type-check valid programs. For rejecting all invalid programs, we employ additional checks when inspecting the typed AST during macro expansion. Over-approximating type correctness in the type level encoding simplifies the encoding. Such approach is especially beneficial when the checks in the macro code are cheaper in terms of compilation performance than Scala's implicit resolution mechanism, which is the case for our approach.

By moving correctness checks to macro code, we reduced the code for the type level encoding from ∼ 600 lines of code in our initial implementation to ∼ 250 lines of code. Issuing compiler errors from macro code also helps in improving error diagnostics since macro code can inspect the AST to give helpful error messages. Debugging compilation issues due to failing implicit resolution, on the other hand, is difficult because the compiler lacks the necessary domain knowledge to hint at which implicit value should have been resolved, resulting in imprecise error messages.

For our purpose of encoding peer and placement types, the key feature of the host language is an expressive type system. Our embedding is based on Scala's unique combination of type system features, namely abstract type members, type refinements, subtyping and path-dependent types. Scala's syntactic flexibility (e.g., writing `T on P` instead of `on[T, P]`) enables an embedding that sorts well with both the host language and the domain-specific abstractions. We conjecture that a similar encoding is possible in languages with similarly expressive type level programming features – of course relying on the type system features of the host language, which might differ from the Scala features which we use. A Haskell implementation, for example, would neither have to support subtyping, nor could it use such type system feature for the encoding. Any domain-specific embedding always compromises between the domain language and host language characteristics to foster easy integration of orthogonal language features of an existing general purpose language and enabling reuse of existing code.

## 5  Macro Expansion

Compiling multitier programs requires splitting the code into deployable components. In ScalaLoci, peer types define the set of components and placement types of value definitions indicate the components to which the values belong. We use Scala macro annotations to split the code by transforming the AST of a multitier module. Scala macro annotations only expand locally, i.e., they only allow the transformation of the annotated class, trait or object. By local expansion, we retain the same support for separate compilation that Scala offers, enabling modular development of multitier applications. Inside every multitier module (i.e., the annotated class, trait or object), we create a nested trait for every peer type which contains the values placed on the peer. This process automatically conflates the values placed on the same peer without requiring the developer to do so manually, disentangling language-level support for modularization form distribution concerns.

**Implementing a Language for Distributed Systems**

To the best of our knowledge, our approach to code splitting is the most extensive use of Scala macros to date, amounting to ∼ 5.5 K lines of code. Our implementation confirms that macro-based code generation is a powerful tool for embedding domain-specific abstractions into Scala using compile-time metaprogramming. Crucially, macros run as part of the Scala compiler and have access to type information of the AST, which is important for our use case of splitting code based on peer types.

Scala supports macros in two different flavors, *def macros* expanding expressions and *macro annotations* expanding declarations of classes, traits, objects or members. Hence, to mark a class, trait or object as multitier module, we rely on macro annotations. In contrast to def macros, which expand typed ASTs, macro annotations expand untyped ASTs. AST transformation before type-checking is considered too powerful since it may change Scala's language semantics significantly [8]. In spirit of keeping our semantics as close as possible to plain Scala, multitier code is syntactically valid and type-correct Scala. Hence, before performing the splitting transformation, we invoke the Scala type checker to obtain a typed AST of the multitier module. Manually invoking the type checker is quite delicate in Scala's current macro system (section 5.5).

Our approach allows accessing libraries from the Java, Scala and Scala.js ecosystems from multitier code (cf. design principle #5), even using mixed Scala/Scala.js multitier modules (section 3.1).

### 5.1 Macros Architecture

Figure 2 provides an overview of the overall architecture of our macro implementation. The Multitier object (at the top) defines the entry point for the macro expansion that is invoked by the Scala compiler to expand @multitier macro annotations. The compiler passes the AST of the annotated module to the macro expansion and retrieves the transformed AST as result. We first run a sequence of preprocessing steps (left side) on the untyped AST, compensating for the lack of contextual abstractions in current Scala (cf. section 5.4). Second, we load a set of components (right side) that constitute the actual code generation. Every component defines potentially multiple processing phases, which specify constraints on whether they should run before/after other phases. All phases run sequentially (satisfying their constraints) to transform certain aspects of the AST. A phase involves one or more AST traversals. So far, we did not optimize the code for minimizing traversals to increase compilation performance.

The processing pipeline first splits the multitier module into its top-level definitions (i.e., the members of the annotated class, trait or object), containing the respective sub-AST together with meta information, such as the peer on which a value is placed extracted from its type. The following phases work on this set of definitions. Instead of using a fully-fledged own intermediate representation, we use standard Scala ASTs enriched with additional information, which proved effective for our use case. The final phase assembles the AST of the complete expanded multitier module.

The largest part of the code base deals with the splitting of placed values (Values component, section 5.2) and the rewriting of remote accesses from direct style via asLocal into calls into the communication backed (RemoteAccess component, section 5.3), which accounts to almost 2 K lines of code.

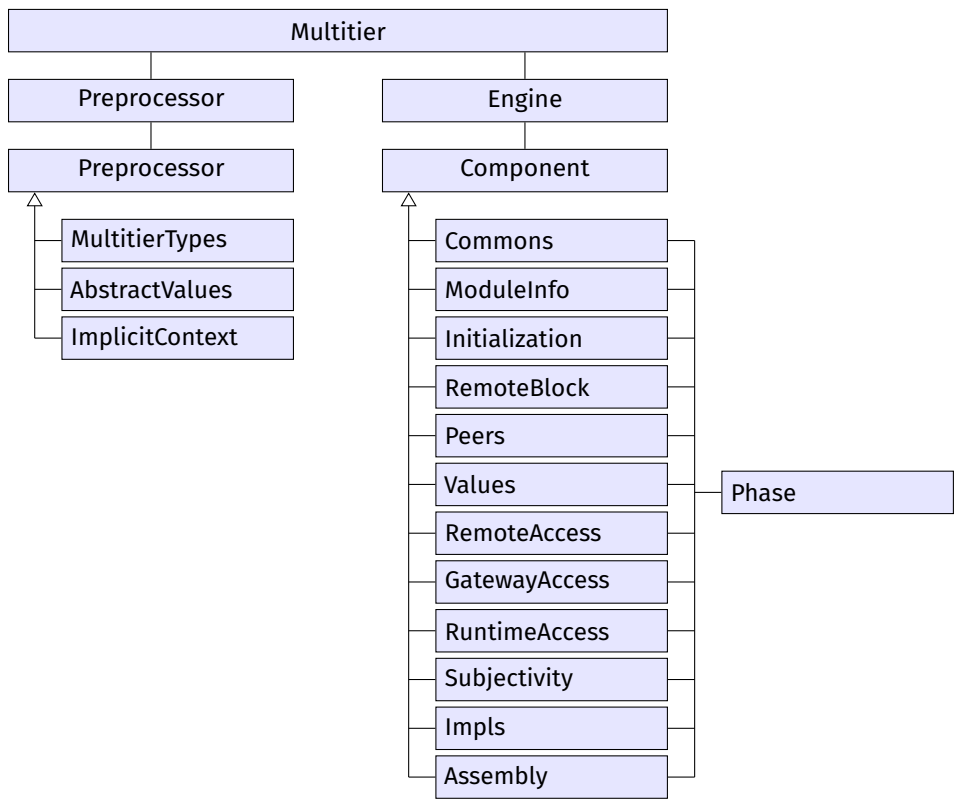

**Figure 2** Software Architecture of the Macro Expansion

### 5.2 Code Splitting Process

The code generation process splits the code according to the placement type of values and creates the necessary run time information for dispatching remote accesses correctly. We consider the following multitier module, which simply defines a single peer MyPeer and a single value i placed on MyPeer:

```
1  @multitier trait SimpleModule {
2    @peer type MyPeer <: { type Tie <: Single[MyPeer] }
3    val i: Int on MyPeer = placed { 1 }
4  }
```

The macro generates signature values to uniquely identify the multitier module and the peer types it defines and creates a runtime representation of the tie specification:

```
1  @MultitierModule trait SimpleModule { /* … */
2    lazy protected val $loci$mod = "SimpleModule"
3    lazy protected val $loci$sig = Module.Signature($loci$mod)
4    lazy val $loci$peer$sig$MyPeer = Peer.Signature("MyPeer", $loci$sig)
5    val $loci$peer$ties$MyPeer = Map($loci$peer$sig$MyPeer -> Peer.Tie.Single)
6  }
```

The signatures and tie specification are used when setting up connections between peer instances at run time to ensure that the connections conform to the static tie constraints. Further, a Marshallable instance – for marshalling and unmarshalling values for network transmission – and a signature for every value is created:

```
@MultitierModule trait SimpleModule { /* ... */
  @MarshallableInfo[Int](0xD89CCAED)
  final protected val $loci$mar$SimpleModule$0 = Marshallable[Int]

  @PlacedValueInfo("i:scala.Int", null, $loci$mar$SimpleModule$0)
  final val $loci$val$SimpleModule$0 = new PlacedValue[Unit, Future[Int]](
    Value.Signature("i:scala.Int", $loci$mod, $loci$sig.path),
    Marshallables.unit, $loci$mar$SimpleModule$0)
}
```

Line 3 resolves a Marshallable instance using Scala's implicit resolution, i.e., it is guaranteed at compile-time that a value is serializable and can be accessed over the network. Line 6 defines the run time representation for the placement of value i, whose remote access does not take any arguments (type Unit) and returns a future (type Future[Int]). Line 7 defines the signature of i for remote dispatch. Line 8 defines the Marshallable instances for the placed value's arguments and its return value.

Marshallable instances require a concrete type, for which the concrete serialization format is known at compile time. Since such information is not available for abstract types – e.g., generic type parameters for parameterized modules – the macro expansion defers the Marshallable resolution to the specific implementations of the multitier module that define a concrete type for the parameter. The following example shows the Marshallable instance (line 2) for a value of type T in a parameterized module (line 1), which delegates to a method (line 5) that is to be implemented in a sub-module:

```
@MultitierModule trait SomeModule[T] { /* ... */
  final protected val $loci$mar$SomeModule$0 = $loci$mar$deferred$SomeModule$0

  @MarshallableInfo[T](0)
  protected def $loci$mar$deferred$SomeModule$0: Marshallable[T, T, Future[T]]
}
```

As a next step, the macro performs the splitting of placed values:

```
@MultitierModule trait SimpleModule { /* ... */
  @compileTimeOnly("Remote access must be explicit.") @MultitierStub
  val i: Int on MyPeer = null.asInstanceOf[Int on MyPeer]

  trait `<placed values of SimpleModule>` extends PlacedValues {
    val i: Int = $loci$expr$SimpleModule$0()
    protected def $loci$expr$SimpleModule$0(): Int = null.asInstanceOf[Int]
  }
  trait $loci$peer$MyPeer extends `<placed values of SimpleModule>` {
    protected def $loci$expr$SimpleModule$0(): Int = 1
  }
}
```

After expansion, the placed value i at the module-level is nulled and annotated to be *compile-time-only* (line 3), i.e., the value cannot be accessed in plain Scala code. The value is kept as a compile-time-only value such that other multitier modules can be type-checked against this module. After type-checking, the macro removes references to compile-time-only values in multitier code. The compile-time-only approach allows us to keep the static specification of placed values (and their placement types) but remove their implementation. Instead, our transformation creates a nested trait for every peer to separate the peer-specific implementations of placed values.

The code generation creates a <placed values> trait nested inside the multitier module (line 5). The trait contains all values of the multitier module. In particular, it defines the placed value i (line 6), which is of type Int (instead of type Int on MyPeer as the module-level definition), i.e., placement types are erased from generated code on the implementation side. Note that placement types are retained on the specification side (line 3) for type-checking at compile time. The value i is initialized by calling the generated method $loci$expr$SimpleModule$0 (line 6), which is nulled by default (line 7). The value is nulled since it is not available on every peer but we need to keep the value in the <placed values> trait to retain Scala's evaluation order. The MyPeer-specific $loci$peer$MyPeer trait specializes the <placed values> trait making for i being initialized to 1 for MyPeer peer instances (line 10).

Finally, for every generated peer trait, the macro synthesizes a $loci$dispatch method, which accesses the placed values for remote accesses:

```
@MultitierModule trait SimpleModule { /* … */
  trait $loci$peer$MyPeer extends `<placed values of SimpleModule>`{ /* … */
    def $loci$dispatch(request: MessageBuffer, signature: Value.Signature,
        reference: Value.Reference) =
      signature match {
        case $loci$val$SimpleModule$0.signature =>
          Try { i } map { response =>
            $loci$mar$SimpleModule$0.marshal(response, reference)
          }
        case _ => super.$loci$dispatch(request, signature, reference)
      }
  }
}
```

For providing a placed value to a remote instance (line 6), the local value is accessed (line 7), potentially unmarshalling its arguments and marshalling its return value (line 8). In case the module does not contain a definition for the value, remote dispatch is delegated to the super module (line 10), mirroring Scala's method dispatch.

### 5.3 Macro Expansion for Remote Access

Since, in our example, we define a MyPeer-to-MyPeer tie, we can access the value i remotely from another MyPeer instance. We add the remote access i.asLocal (line 4):

```
@multitier trait SimpleModule {
  @peer type MyPeer <: { type Tie <: Single[MyPeer] }
  val i: Int on MyPeer = placed { 1 }
  val j: Future[Int] on MyPeer = placed { i.asLocal }
}
```

Similar to the expansion of value i (cf. section 5.2), the definition of value j is extracted into a peer-specific method $loci$expr$SimpleModule$1:

```
1  @MultitierModule trait SimpleModule { /* … */
2    trait $loci$peer$MyPeer extends `<placed values of SimpleModule>`{ /* … */
3      protected def $loci$expr$SimpleModule$1(): Unit =
4        BasicSingleAccessor[Int, MyPeer, Future[Int], MyPeer](RemoteValue)(
5          new RemoteRequest(
6            (), $loci$val$SimpleModule$0, $loci$peer$sig$MyPeer, $loci$sys)
7          ).asLocal
8    }
9  }
```

The transformation from the i.asLocal user code to the call into the runtime system (lines 4 to 7) ties the knot between the direct-style remote access of ScalaLoci multitier code and the message-passing-based network communication of our communication backend (section 6). The interface for remote access (i.e., the asLocal call in the example) is declared by an implicit class. In the example, the interface is defined by the BasicSingleAccessor implicit class, which requires an implicit Transmission argument for accessing the remote value (cf. section 4.2). The Transmission argument is rewritten by the macro to a RemoteRequest (line 5) that is instantiated with (i) the arguments for the remote call, (ii) the signature of the accessed value, (iii) the signature of the peer on which the value is placed and (iv) a reference to the runtime system (inherited from PlacedValues trait) that manages the network connections. With these information assembled by macro expansion, asLocal can perform the remote access.

### 5.4 Peer Contexts

Before invoking the type checker, the macro performs a transformation step on the untyped AST to compensate for the lack of contextual abstractions in Scala 2, which are to be available Scala 3 [39]. The current context determines for any expression to which peer it belongs (cf. section 4.1). Since the context needs to be available to the type checker, the transformation has to take place before type-checking. It transforms placed expressions of the form placed { e } to placed { implicit ! => e }, where ! is the name of the argument carrying the (implicit) peer context. In the lexical scope of the expression e, the context can be resolved by the compiler from the implicit scope. For better IDE support, the implicit argument can also be written explicitly by the developer, in which case we do not transform the expression.

### 5.5 Interaction with the Type System

Since (i) we rely on type-checked ASTs for guiding the code splitting by placement types and (ii) splitting changes the shape of the multitier module (i.e., adding members to the annotated module), essentially changing the module's type, the AST transformation needs to be performed during compilation. Scala's macro system enables such interaction with the type system, which is essential for splitting ScalaLoci multitier code, in contrast to code generation approaches that run strictly before the compiler.

Yet, in our experience, invoking the type checker for annotated classes, traits or objects is quite fragile with the current Scala macro system. Type-checking the AST again after transformation, where trees are re-type-checked in a different lexical context after transformation, can easily corrupt the owner chain of the compiler's symbol table. To work around those issues, we implemented a transformation that converts ASTs such that they can be re-type-checked. This transformation is independent of the code splitting and is available as a separate project.[3]

Type-checking multitier modules expands all nested macro invocations. We extensively used `ScalaLoci` with a domain-specific language for reactive programming that relies on *def macros* (i.e., expression-based macros). We did not observe any issue with mixing different macro-based language extensions. Invoking the type checker for *macro annotations* (i.e., annotation-based macros) on modules which are themselves nested into other modules, however, is not supported by the current macro system.

**5.6 Lessons Learned**

In our experience, performing complex AST transformations is quite involved using the current Scala macro system, which lacks built-in support for automatic hygiene [9], i.e., separating the lexical scopes of the macro implementation and the macro call site to guarantee the absence of name clashes between user code and code generated by macro expansion. The macro developer is responsible for ensuring that the generated code does not interfere with the lexical scope of the macro call site by creating identifier names that are expected to be unique or using fully qualified names. Moreover, the macro system exposes compiler internals such as the compiler's symbol table, which developers have to keep consistent when transforming typed ASTs. When moving ASTs between different contexts or mixing typed ASTs with newly generated untyped ASTs, developers have to fix the symbol chain manually or re-type-check the AST.

This complex interaction with the type system is the reason why the macro system considered for the next version of Scala (and currently being implemented in the Dotty compiler) does not allow explicit interaction with the type system [40]. The new TASTy reflection API properly abstracts over compiler internals and only supports ASTs that are already typed. Macro systems of other languages, such as Racket, are more powerful, supporting the addition of new syntactic forms through the transformation of arbitrary Racket syntax. The revised Scala macros, however, are still more powerful than macro systems like Template Haskell, which do not take the AST as input without explicitly quoting expressions at the call site. Expanding macros on typed ASTs helps in controlling the power of macros and keeping syntactic and semantic deviations from plain Scala small. Prohibiting any kind of interaction with the type system, however, seems too limiting and would make a macro-based implementation of `ScalaLoci` impossible. Other use cases currently covered by macro annotations are also not well supported under the new scheme, e.g., auto-generation of lenses for manipulating data structures [57] or auto-generation of serializers [5]. To restore support for such use cases, we could imagine an approach that allows macros to change types in a clearly-defined and controlled way. For instance, the macro could be expanded in several phases that allow for different kinds of modifications:

1. In a first phase, the macro can inspect (but not transform) the current untyped AST and declare members, super traits or super classes and annotations that should be added to the annotated class, trait, object or its companion. Only declarations are required for the following type-checking, not necessarily their definitions.
2. The complete code is type-checked.
3. Similar to the first phase, the macro can inspect (but not transform) the tree which, in contrast to the first phase, is now type-checked. The macro can again declare members, super traits or super classes and annotations that should be added to the annotated class, trait, object or its companion. It may be necessary to further restrict the members, which could be declared, e.g., disallowing adding members that interfere with implicit or method overload resolution since both already happened as part of the type-checking in the second phase.
4. The new member declarations are type-checked. Since no members can be removed and adding members can be restricted appropriately, it is sufficient to only type-check the new members.
5. Finally, macro annotations are expanded. Macro expansion works on type-checked ASTs. Members generated in the previous phases are visible to the macro.

---

[3] http://github.com/stg-tud/retypecheck

We believe that well-defined interfaces for macros are essential to retain the current level of usefulness of macro annotations in a future macro system while avoiding the issues of the current macro system.

## 6  Runtime

The ScalaLoci communication runtime hides the implementation details of network communication, e.g., data serialization and underlying network protocols, such that developers can access remote values in direct style (via asLocal) instead of explicitly sending network messages and registering callbacks for receiving messages. Figure 3 shows the communication runtime which underlies a ScalaLoci multitier program. Our runtime system provides abstraction layers for different network protocols, serialization schemes and the type-safe transmission of values.

### 6.1 Communicators

The lower layer defines *communicators* abstracting over network protocols. We currently support TCP (on the JVM), WebSocket (on the JVM and in web browsers) and WebRTC (in web browsers). Communicators can be instantiated in *listening* mode (e.g., binding a local TCP port and listening for incoming connections) or *connecting* mode (e.g., initiating a TCP connection to a remote host). After establishing a connection, the communicators of both endpoints create a Connection object that provides a bidirectional message-passing channel, abstracting over the communication model of the underlying protocol, such as TCP byte streams or WebSocket messages. Communicators also offer additional meta information about the established network connection to the higher level, such as if the connection is secure (i.e., encrypted and integrity-protected) or authenticated (and which user token or certificate was used for authentication). Currently, communicators are restricted to point-to-point bidirectional channels. Yet, in the future, we may support additional communication schemes, e.g., broadcasting, single-shot request–response, etc.

**6.2 Serializers**

To serialize values of a specific type for network transmission, the runtime requires an implementation of the Serializable type class for every such type – encoded in Scala using the *concept pattern* [41]. The compiler derives Serializable instances using Scala's implicit resolution, guaranteeing that values of a certain type are serializable.

The type class Serializable[T] witnesses that a value of type T is serializable by providing methods for both serialization and deserialization:

```
1  trait Serializable[T] {
2    def serialize(value: T): MessageBuffer
3    def deserialize(value: MessageBuffer): Try[T]
4  }
```

The runtime invokes the Serializable methods to convert between a value of type T and a MessageBuffer, buffering the byte array to be sent or received over the network. Serialization is not expected to fail, but deserialization may result in a runtime error if the buffer does not contain a valid serialization. Hence, the return value is wrapped in a Try, which represents either a success value or a failure.

We implemented two serializers that simply forward to the µPickle [27] or circe [5] serialization libraries, respectively. The µPickle serializer, for example, is declared as an implicit value of type Serializable[T] given that the compiler is able to resolve implicit instances for Writer[T] and Reader[T] (which are type classes defined by µPickle for serializing and deserializing values):

```
1  implicit def upickleSerializable[T]
2    (implicit writer: Writer[T], reader: Reader[T]): Serializable[T] = // ...
```

For a required implicit Serializable[Int] instance, the compiler automatically resolves the call upickleSerializable(upickle.default.IntWriter, upickle.default.IntReader), constructing a Serializable[Int] instance based on µPickle's IntWriter and IntReader.

**6.3 Transmitters**

The higher level defines *transmitters* implementing the transmission semantics specific to a certain data type. To make a type of value available for transmission, the runtime requires an implementation of the Transmittable type class for every such type. A Transmittable[B, I, R] instance witnesses that a value of type T can be send over the network as value of type I and whose local representation after remote access is of

**Implementing a Language for Distributed Systems**

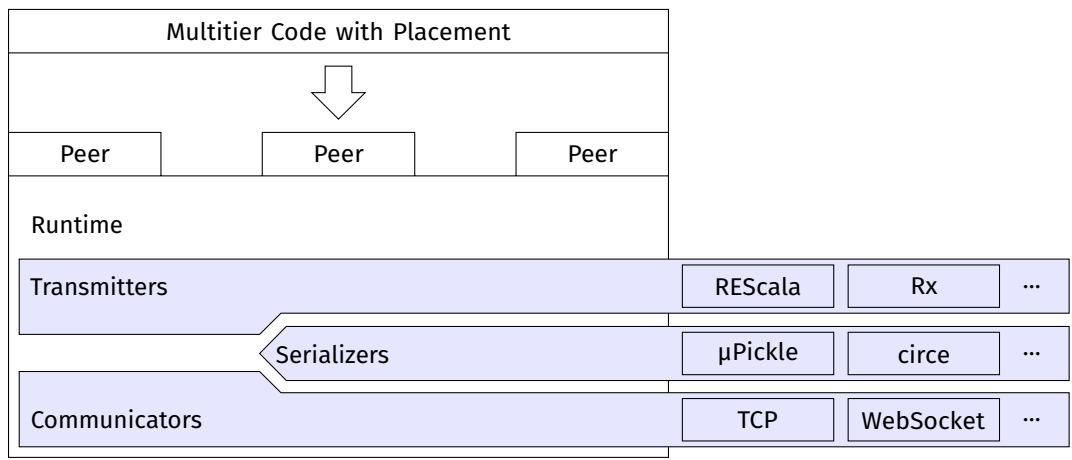

**Figure 3** Communication Runtime

type R. When accessing a value remotely, the runtime creates related Transmittable instances on both connection endpoints.

Primitive and standard collection values are retrieved from a remote instance in a pull-based fashion upon each request. However, accessing event streams, on the other hand, does not exhibit pull-based semantics. Instead, events are pushed to remote instances for every event occurrence. Runtime support for accessing event streams remotely is crucial since communication in distributed systems is often event-based [11, 31] and event streams allow for data across hosts to be specified in a declarative way (cf. design principle #4). To support such use case, the runtime allows transmitters to operate in *connected* mode, providing a typed message channel between both communication endpoints. The remote event stream uses the channel to propagate events in a push-based manner over the network. The runtime can multiplex multiple such message channels over the same underlying network connection.

The runtime comes with built-in support for transmitting primitive values and standard collections. We further implemented transmitters for REScala [50] reactives and Rx [32] observables, which developers can plug in when needed. The runtime is extensible by defining Transmittable type class instances for additional types.

Transmitters abstract over different semantics for propagating values to remote hosts. Depending on the type of the value that is accessed remotely, the compiler automatically selects an appropriate transmitter through implicit resolution. The communication runtime performs the actual network communication for remote accesses at the ScalaLoci language level that are transformed into calls into the runtime during macro expansion (cf. section 5.3). The communication runtime can also be used independently of the macro-based language implementation to abstract over transmission semantics, serialization and network protocols.

### 6.4 Lessons Learned

Instances of the Transmittable type class, which implement the transmission semantics for a specific type, can be nested, e.g., transmitting an *n*-tuple requires a Transmittable instance for every of the *n* elements. For accessing a value of type `T` remotely, the compiler has to resolve both a `Transmittable` instance and a serializer from the implicit scope. In our experience, failing to resolve a (deeply) nested implicit value leads to less-than-optimal error messages by the compiler because the compiler only reports on the outermost implicit that could not be resolved since it is generally not clear for which of the possibly multiple alternatives for resolving nested implicit values a developer expected implicit resolution to succeed. In our use case, we expect that `Transmittable` instances should always be resolvable or the compiler should issue an error if no matching `Transmittable` instance is in scope.

For such use cases, we propose the following scheme to achieve more precise error messages: We provide a fallback implicit value for `Transmittable`, which is defined in the `Transmittable` object. Such values are resolved by the Scala compiler with lower priority in case there is no other matching implicit value in scope. This fallback value can always be resolved but resolution results in a reference to a *compile-time-only* value, i.e., carrying the `compileTimeOnly` annotation. With this scheme, implicit resolution for `Transmittable` instances always succeeds. If there is no usable `Transmittable` instance in scope, however, the resolved fallback value results in a more meaningful compiler error that hints at the nested `Transmittable` which could not be resolved.

Another issue with Scala implicits we encountered is their lack of global coherence guarantees. In contrast to Haskell, where type class instances are globally unique, i.e., there is at most one type class implementation for every type in the whole program, Scala allows different type class instances for the same type. Precedence rules for implicit resolution decide which type class instance is chosen by the compiler. Coherence is important for our use case, since the definition site of a placed value – and the generated dispatch logic for remote accesses (section 5.2) – and the remote call site of a placed value (section 5.3) need to agree on the transmission semantics implemented by the `Transmittable` type class instance. By inspecting the AST, containing the values implicitly resolved by the compiler, during macro expansion, we ensure that `Transmittable` instances are coherent or issue a compiler error otherwise.

## 7 Related Work

**Multitier Programming** Multitier programming [13, 14, 43, 45, 46, 47, 53] is in the tradition of programming languages for distributed systems with influential languages such as Argus [28], Emerald [4], Distributed Oz [22, 58] and Dist-Orc [2]. More recent contributions focus on specific design aspects, e.g., cloud types to ensure eventual consistency [7], conflict-free replicated data types (CRDT) [55], language support for safe distribution of computations [34] and fault tolerance [33].

Multitier languages emerge in the web context to remove the separation between client and server code, either by compiling the client side to JavaScript or by adopting JavaScript for the server, too. Hop [53] and Hop.js [54] are dynamically typed languages that follow a traditional client–server communication scheme with asynchronous callbacks. In Links [14, 18] and Opa [46], functions are annotated to specify either client- or server-side execution. Both languages also follow the client–server model and feature a static type system. In StiP.js [43, 44], annotations assign code fragments to the client or the server. Slicing detects the dependencies between each fragment and the rest of the program. In contrast, in ScalaLoci, developers specify placement *in types*, enabling architectural reasoning. All approaches above focus on the web, contrarily to our goal of supporting other architectures. An exception is ML5 [37]: *Possible worlds*, as known from modal logic, address the purpose of placing computations and, similar to ScalaLoci, are part of the type. ML5, however, does not support architecture specification, i.e., it does not allow for expressing different architectures in the language and was only applied to the Web setting so far.

**Metaprogramming** Compile-time metaprogramming was pioneered by the Lisp macro system [23] that supports transformations of arbitrary Lisp syntax, facilitating the addition of new syntactic forms. Racket, a Lisp dialect, is designed to allow building new languages based on macros [17]. In contrast, but in line with Scala's upcoming macro system that only supports typed ASTs, the main part of our macro transformation is not a pure syntax-to-syntax transformation, but works on ASTs that are already typed. Thus, Scala macros are more restrictive than Lisp macros since untyped Scala macros are deemed too powerful [8], e.g., hindering tool support.

In Template Haskell [56], Haskell metaprograms generate ASTs, which the compiler splices into the call sites. Such metaprograms do not take ASTs as input without explicitly using quasiquotation at the call site. Template Haskell has been used to optimize embedded domain-specific languages at compile time [52].

Rust supports hygienic *declarative macros* that expand before type-checking and define rewrite rules to transform programs based on syntactic patterns. Rust's *procedural macros* are more powerful using Rust code to rewrite token streams and accessing compiler APIs. A combination of Rust's type system and macro system was shown to support a shallow embedding of the lambda calculus [24].

**Related Paradigms** The actor model [25] encapsulates control and state into concurrent units that exchange asynchronous messages. The resulting decoupling by asynchronous communication and by the no-shared-memory approach enables scalability and fault tolerance.

Reactive programming [3] provides abstractions for defining time-changing values (signals), event streams and their combination. In line with multitier programming, distributed reactive programming [35, 36, 29] allows developers to define event streams that span over different machines.

Software architectures [19, 42] organize software systems into components and their connections as well as constraints on their interaction. Architecture description languages (ADL) [30] provide a mechanism for high-level specification and analysis of large software systems, for example, to guide architecture evolution. Partitioned Global Address Space Languages (PGAS), such as X10 [15], provide a programming model for high-performance parallel execution. PGAS languages define a globally shared address space aiming at a goal similar to multitier languages – reduce boundaries among hosts.

## 8 Conclusion

In this paper, we presented the implementation of the ScalaLoci multitier language and its design as a domain-specific language embedded into Scala. We show how to exploit Scala's advanced language features, i.e, type level and macro programming, for embedding ScalaLoci abstractions. We reported on our experiences with such an approach and the challenges we needed to tackle and how they led to the current ScalaLoci architecture and implementation strategy.

**Acknowledgements** This work has been co-funded by the Deutsche Forschungsgemeinschaft (DFG, German Research Foundation) – SFB 1053 – 210487104 – and – SFB 1119 – 236615297, by the DFG projects 322196540 and 383964710, by the LOEWE initiative (Hesse, Germany) within the emergenCITY centre and within the Software-Factory 4.0 project and by the German Federal Ministry of Education and Research and the Hessian State Ministry for Higher Education, Research and the Arts within their joint support of the National Research Center for Applied Cybersecurity ATHENE.

## About the authors

**Pascal Weisenburger** is a PhD student at the Technical University of Darmstadt. His research interests focus on programming language design, in particular multitier programming, reactive programming and event-based systems. He is the main developer of the ScalaLoci multitier programming language. Contact him at weisenburger@cs.tu-darmstadt.de.

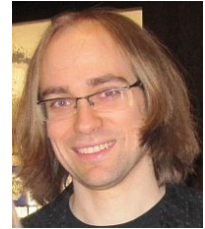

**Guido Salvaneschi** is an assistant professor at TU Darmstadt. He completed his PhD in the Dipartimento di Elettronica e Informazione at Politecnico di Milano, under the supervision of Prof. Carlo Ghezzi, with a doctoral dissertation on language-level techniques for adaptive software. His research interests focus on programming language design and software engineering, in particular for distributed systems and reactive/event-based applications. His most recent publications appeared at OOPSLA, ECOOP, PLDI, ICFP, ICSE, FSE, TSE and ASE. Contact him at salvaneschi@cs.tu-darmstadt.de.

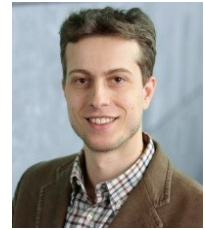